\PassOptionsToPackage{unicode}{hyperref}
\PassOptionsToPackage{hyphens}{url}
\PassOptionsToPackage{dvipsnames,svgnames,x11names}{xcolor}
\documentclass[
]{article}
\usepackage{amsmath,amssymb}
\usepackage{lmodern}
\usepackage{iftex}
\ifPDFTeX
  \usepackage[T1]{fontenc}
  \usepackage[utf8]{inputenc}
  \usepackage{textcomp} % provide euro and other symbols
\else % if luatex or xetex
  \usepackage{unicode-math}
  \defaultfontfeatures{Scale=MatchLowercase}
  \defaultfontfeatures[\rmfamily]{Ligatures=TeX,Scale=1}
\fi
\IfFileExists{upquote.sty}{\usepackage{upquote}}{}
\IfFileExists{microtype.sty}{% use microtype if available
  \usepackage[]{microtype}
  \UseMicrotypeSet[protrusion]{basicmath} % disable protrusion for tt fonts
}{}
\makeatletter
\@ifundefined{KOMAClassName}{% if non-KOMA class
  \IfFileExists{parskip.sty}{%
    \usepackage{parskip}
  }{% else
    \setlength{\parindent}{0pt}
    \setlength{\parskip}{6pt plus 2pt minus 1pt}}
}{% if KOMA class
  \KOMAoptions{parskip=half}}
\makeatother
\usepackage{xcolor}
\newlength{\cslhangindent}
\newlength{\csllabelwidth}
\newlength{\cslentryspacingunit} % times entry-spacing
\newenvironment{CSLReferences}[2] % #1 hanging-ident, #2 entry spacing
 {% don't indent paragraphs
  \setlength{\parindent}{0pt}
  \ifodd #1
  \let\oldpar\par
  \def\par{\hangindent=\cslhangindent\oldpar}
  \fi
  \setlength{\parskip}{#2\cslentryspacingunit}
 }%
 {}
\usepackage{calc}

\ifLuaTeX
\usepackage[bidi=basic]{babel}
\else
\usepackage[bidi=default]{babel}
\fi
\babelprovide[main,import]{american}
\def\languageshorthands#1{}
\ifLuaTeX
  \usepackage{selnolig}  % disable illegal ligatures
\fi
\IfFileExists{bookmark.sty}{\usepackage{bookmark}}{\usepackage{hyperref}}
\IfFileExists{xurl.sty}{\usepackage{xurl}}{} % add URL line breaks if available
\hypersetup{
  pdftitle={SERENE: The Semi-Automatic User Experience Detector},
  pdfauthor={Andrea Esposito},
  pdflang={en-US},
  colorlinks=true,
  linkcolor={Maroon},
  filecolor={Maroon},
  citecolor={Blue},
  urlcolor={Blue},
  pdfcreator={LaTeX via pandoc}}

\title{SERENE: The Semi-Automatic User Experience Detector}

\usepackage[affil-it]{authblk}
\usepackage{orcidlink}
\author[1%
  ]{Andrea Esposito%
    \,\orcidlink{0000-0002-9536-3087}\,\\
    \href{mailto:andrea.esposito@uniba.it}{\texttt{andrea.esposito@uniba.it}}
    }

\affil[1]{Department of Computer Science, University of Bari Aldo Moro,
Bari, Italy}
\date{}

\begin{document}
\maketitle

\hypertarget{summary}{%
\section{Summary}\label{summary}}

SERENE (uSer ExpeRiENce dEtector), also known as UX-SAD (User
eXperience-Smells Automatic Detector), is a research project born in
2020, which comprises different components. As its name suggests, its
primary goal is to provide a way to quickly and (semi-) automatically
detect problems in the user experience of websites and web-based
systems. Through a set of Artificial Intelligence (AI) models, SERENE
detects users' emotions in web pages while guaranteeing users' privacy.
Its main strength over typical user experience and usability evaluation
is in the generalizability of its detections. While traditional methods
use samples (that may not be representative), SERENE allows to tap into
data provided by the whole user population. The platform is available at
\url{https://serene.ddns.net}.

\hypertarget{statement-of-need}{%
\section{Statement of need}\label{statement-of-need}}

Despite the well-documented benefits of usability evaluation methods,
they are often neglected by many companies and practitioners. This is
primarily due to the perception that usability experts are scarce
(\protect\hyperlink{ref-Vanderdonckt2004Automated}{Vanderdonckt et al.,
2004}), and that these methods require significant resources that may
not be well-suited to their needs
(\protect\hyperlink{ref-Ardito2014Investigating}{Ardito et al., 2014}).
However, it is widely recognized that incorporating usability
evaluations can significantly enhance the overall quality of products
(\protect\hyperlink{ref-Dingli2011USEFul}{Dingli \& Mifsud, 2011}). To
overcome these challenges, automatic or semi-automatic tools can be
employed to assist evaluators with insufficient skills in conducting
reliable usability evaluations. By utilizing these tools, usability
evaluations can be made more efficient, and tailored to better address
the specific needs of companies.

User eXperience (UX) has become an increasingly important aspect of
software. It is defined by (\protect\hyperlink{ref-ISO2018924111}{ISO,
2018}) as ``a person's perceptions and responses resulting from the use
and/or anticipated use of a product, system, or service.'' Therefore, in
general, designing for UX is more than designing for the traditional
attributes of usability, as it also focuses on the hedonic aspects of
the interaction (\protect\hyperlink{ref-Law2009Understanding}{Law et
al., 2009}). Since emotions are important elements of UX, some authors
have been looking for ways of identifying users' emotions by analyzing
their interaction with systems
(\protect\hyperlink{ref-Desolda2021Detecting}{Desolda et al., 2021}).

SERENE is a web platform designed for UX experts to aid the UX
evaluation of websites
(\protect\hyperlink{ref-Esposito2022SERENE}{Esposito et al., 2022}). In
particular, evaluators are guided in the discovery of ``UX Smells''
(\protect\hyperlink{ref-Buono2020Detection}{Buono et al., 2020})
employing heatmaps, that show the concentration of emotions in the
webpage. The ground assumption of this methodology is that areas of the
page with usability or UX problems evoke negative emotions in their
users (\protect\hyperlink{ref-Li2018Effects}{Li et al., 2018}).

\hypertarget{research-informed-design}{%
\section{Research-Informed Design}\label{research-informed-design}}

The design of SERENE derives from user research, following a typical
human-centered design (HCD) approach
(\protect\hyperlink{ref-ISO20199241210}{ISO, 2019}). This section
briefly details the design of the various components of SERENE.

\hypertarget{emotion-detection-models}{%
\subsection{Emotion Detection Models}\label{emotion-detection-models}}

Multiple steps were required to build the privacy-conscious emotion
detection component. The initial step was the collection of an
in-the-wild dataset of interaction logs (i.e., mouse movements and
clicks, as well as aggregated keyboard usage data) linked to emotions
(\protect\hyperlink{ref-Desolda2021Detecting}{Desolda et al., 2021}).
The emotions were collected through facial recognition using
state-of-the-art techniques
(\protect\hyperlink{ref-Desolda2021Detecting}{Desolda et al., 2021}).
Following the dataset collection phase, multiple machine-learning models
were compared to select the better-performing ones for each emotion
(\protect\hyperlink{ref-Desolda2021Detecting}{Desolda et al., 2021}).

\hypertarget{visualization-of-usability-issues}{%
\subsection{Visualization of Usability
Issues}\label{visualization-of-usability-issues}}

Various approaches are available to visualize the automatically detected
usability issues. Following the automation level framework proposed by
Parasuraman et al. (\protect\hyperlink{ref-Parasuraman2000Model}{2000}),
then evolved by Shneiderman
(\protect\hyperlink{ref-Shneiderman2020HumanCentered}{2020}), three
different visualizations were designed. Namely, the classification
output can be presented using a full automation, a full control, or a
middle-ground solution. Through a user study, Esposito et al.
(\protect\hyperlink{ref-Esposito2024Fine}{2024}) highlighted that a full
control or a middle-ground solution allows for the discovery of a higher
amount of usability issues. Therefore a full control solution is
implemented in this version of the platform, although future versions
may allow users to select their preferred visualization style (depending
on their goal).

\hypertarget{acknowledgments}{%
\section{Acknowledgments}\label{acknowledgments}}

The research of Andrea Esposito is funded by a Ph.D.~fellowship within
the framework of the Italian ``D.M. n.~352, April 9, 2022g - under the
National Recovery and Resilience Plan, Mission 4, Component 2,
Investment 3.3 - Ph.D.~Project''Human-Centered Artificial Intelligence
(HCAI) techniques for supporting end users interacting with AI
systems'', co-supported by ``Eusoft S.r.l.'' (CUP H91I22000410007).
Andrea Esposito acknowledges the help of Giuseppe Desolda (without whom
this idea wouldn't be born) and Rosa Lanzilotti, who both took part in
the HCD of the various components of the system.

\hypertarget{references}{%
\section*{References}\label{references}}
\addcontentsline{toc}{section}{References}

\hypertarget{refs}{}
\begin{CSLReferences}{1}{0}
\leavevmode\vadjust pre{\hypertarget{ref-Ardito2014Investigating}{}}%
Ardito, C., Buono, P., Caivano, D., Costabile, M. F., \& Lanzilotti, R.
(2014). Investigating and promoting {UX} practice in industry: {An}
experimental study. \emph{International Journal of Human-Computer
Studies}, \emph{72}(6), 542--551.
\url{https://doi.org/10.1016/j.ijhcs.2013.10.004}

\leavevmode\vadjust pre{\hypertarget{ref-Buono2020Detection}{}}%
Buono, P., Caivano, D., Costabile, M. F., Desolda, G., \& Lanzilotti, R.
(2020). Towards the detection of {UX} smells: {The} support of
visualizations {[}Journal Article{]}. \emph{IEEE Access : Practical
Innovations, Open Solutions}, \emph{8}, 6901--6914.
\url{https://doi.org/10.1109/access.2019.2961768}

\leavevmode\vadjust pre{\hypertarget{ref-Desolda2021Detecting}{}}%
Desolda, G., Esposito, A., Lanzilotti, R., \& Costabile, M. F. (2021).
Detecting {Emotions Through Machine Learning} for {Automatic UX
Evaluation}. In C. Ardito, R. Lanzilotti, A. Malizia, H. Petrie, A.
Piccinno, G. Desolda, \& K. Inkpen (Eds.), \emph{Human-{Computer
Interaction} -- {INTERACT} 2021} (Vol. 12934, pp. 270--279). Springer
International Publishing.
\url{https://doi.org/10.1007/978-3-030-85613-7_19}

\leavevmode\vadjust pre{\hypertarget{ref-Dingli2011USEFul}{}}%
Dingli, A., \& Mifsud, J. (2011). {USEFul}: {A Framework} to {Mainstream
Web Site Usability} through {Automated Evaluation}. \emph{International
Journal of Human Computer Interaction}, 2011--2010.

\leavevmode\vadjust pre{\hypertarget{ref-Esposito2024Fine}{}}%
Esposito, A., Desolda, G., \& Lanzilotti, R. (2024). The fine line
between automation and augmentation in website usability evaluation.
\emph{Scientific Reports}, \emph{14}(1), 10129.
\url{https://doi.org/10.1038/s41598-024-59616-0}

\leavevmode\vadjust pre{\hypertarget{ref-Esposito2022SERENE}{}}%
Esposito, A., Desolda, G., Lanzilotti, R., \& Costabile, M. F. (2022).
{SERENE}: {A Web Platform} for the {UX Semi-Automatic Evaluation} of
{Website}. \emph{Proceedings of the 2022 International Conference on
Advanced Visual Interfaces}.
\url{https://doi.org/10.1145/3531073.3534464}

\leavevmode\vadjust pre{\hypertarget{ref-ISO2018924111}{}}%
ISO. (2018). \emph{9241-11:2018 {Ergonomics} of human-system interaction
--- {Part} 11: {Usability}: {Definitions} and concepts} (Journal Article
No. 9241-11).

\leavevmode\vadjust pre{\hypertarget{ref-ISO20199241210}{}}%
ISO. (2019). \emph{9241-210:2019 {Ergonomics} of human-system
interaction --- {Part} 210: {Human-centred} design for interactive
systems} (Journal Article No. 9241-210).

\leavevmode\vadjust pre{\hypertarget{ref-Law2009Understanding}{}}%
Law, E. L.-C., Roto, V., Hassenzahl, M., Vermeeren, A. P. O. S., \&
Kort, J. (2009). Understanding, scoping and defining user experience:
{A} survey approach. \emph{Proceedings of the {SIGCHI} Conference on
Human Factors in Computing Systems}, 719--728.
\url{https://doi.org/10.1145/1518701.1518813}

\leavevmode\vadjust pre{\hypertarget{ref-Li2018Effects}{}}%
Li, X., Xiao, Z., \& Cao, B. (2018). Effects of {Usability Problems} on
{User Emotions} in {Human}--{Computer Interaction}. In S. Long \& B. S.
Dhillon (Eds.), \emph{Man--{Machine}--{Environment System Engineering}}
(Vol. 456, pp. 543--552). Springer Singapore.
\url{https://doi.org/10.1007/978-981-10-6232-2_63}

\leavevmode\vadjust pre{\hypertarget{ref-Parasuraman2000Model}{}}%
Parasuraman, R., Sheridan, T. B., \& Wickens, C. D. (2000). A model for
types and levels of human interaction with automation. \emph{IEEE
Transactions on Systems, Man, and Cybernetics - Part A: Systems and
Humans}, \emph{30}(3), 286--297.
\url{https://doi.org/10.1109/3468.844354}

\leavevmode\vadjust pre{\hypertarget{ref-Shneiderman2020HumanCentered}{}}%
Shneiderman, B. (2020). Human-{Centered Artificial Intelligence}:
{Reliable}, {Safe} \& {Trustworthy} {[}Journal Article{]}.
\emph{International Journal of Human--Computer Interaction},
\emph{36}(6), 495--504.
\url{https://doi.org/10.1080/10447318.2020.1741118}

\leavevmode\vadjust pre{\hypertarget{ref-Vanderdonckt2004Automated}{}}%
Vanderdonckt, J., Beirekdar, A., \& Noirhomme-Fraiture, M. (2004).
Automated {Evaluation} of {Web Usability} and {Accessibility} by
{Guideline Review}. In T. Kanade, J. Kittler, J. M. Kleinberg, F.
Mattern, J. C. Mitchell, M. Naor, O. Nierstrasz, C. Pandu Rangan, B.
Steffen, M. Sudan, D. Terzopoulos, D. Tygar, M. Y. Vardi, G. Weikum, N.
Koch, P. Fraternali, \& M. Wirsing (Eds.), \emph{Web {Engineering}}
(Vol. 3140, pp. 17--30). Springer Berlin Heidelberg.
\url{https://doi.org/10.1007/978-3-540-27834-4_4}

\end{CSLReferences}

\end{document}